\documentclass[fleqn,12pt,twoside]{article}
\usepackage{espcrc1}

\usepackage{graphicx}
\usepackage[figuresright]{rotating}

\newcommand{\AmS}{{\protect\the\textfont2
  A\kern-.1667em\lower.5ex\hbox{M}\kern-.125emS}}

\title{Multifragmentation and the liquid-gas phase transition:
an experimental overview}

\author{W. Trautmann\address[GSI]
{Gesellschaft f{\"u}r Schwerionenforschung mbH,\\
Planckstr. 1, D-64291 Darmstadt, Germany}
}
      
\begin{document}

\maketitle

\begin{abstract}Two roads are presently being 
followed in order to establish 
the existence of a liquid-gas phase transition in finite
nuclear systems from nuclear reactions at high energy. 
The clean experiment of observing the thermodynamic properties 
of a finite number of nucleons in a container is presently only 
possible with the computer. Performed with 
advanced nuclear transport models, it has revealed the first-order 
character of the transition and allowed the
extraction of the pertinent thermodynamic parameters. The validity 
of the applied theory is being confirmed by comparing its predictions 
for heavy-ion reactions with exclusive experiments.\\
The second approach is experimentally more direct. 
Signals of the transition are searched for by analysing reaction 
data within the framework of thermodynamics of small systems. 
A variety of potential signals has been investigated 
and found to be qualitatively consistent with the expectations 
for the phase transition. Many of them are 
well reproduced with percolation models which places the nuclear 
fragmentation into the more general context of 
partitioning phenomena in finite systems.\\
A wealth of new data on this subject has been obtained in recent 
experiments, some of them with a new generation 
of multi-detector devices aiming at higher resolutions, isotopic 
identification of the fragments, and the coincident detection of 
neutrons. Isotopic effects in multifragmentation were addressed 
quite intensively, with particular attention being given to their 
relation to the symmetry energy and its dependence on density.

\end{abstract}

\section{INTRODUCTION}

Multifragmentation reactions are often primarily seen and
discussed in the context of the nuclear liquid-gas phase transition. 
Phase transitions in nuclear matter are indeed very special because 
they occur on temperature, pressure, and energy scales many orders of 
magnitude away from those familiar from ordinary matter. 
It is an interesting 
question in what form existing concepts for phase transitions can 
be applied to the phenomena governed by the strong force. 
The observation or proof of the existence of a phase transition has 
been an important objective of heavy-ion physics for quite some time.

There is little doubt that a liquid-gas phase transition should
exist in extended nuclear matter \cite{sauer76,jaqa83,muell95}. 
It is also known that the corresponding 
conditions of density and temperature can be reached in reactions 
between finite nuclei. The problems encountered there, however, 
are not only related to the small size of the reaction system, 
i.e. to the severely limited number of constituents, but also to 
possible dynamical effects which may hinder the homogeneous 
population of the phase space. A third difficulty arises from 
the limits in our control of the experiments. 
The identification of the complex reaction processes is incomplete, 
and even the detected fraction of a multi-particle event is only 
known with finite precision. To sort data according to impact 
parameter, excitation energy or temperature is limited in accuracy 
by statistical and systematic errors.
To work with the best possible coverage and resolution is thus 
even more essential, which has motivated a continuing process of 
improving the experimental possibilities and devices.

More recently, a new direction for studying multifragmentation 
reactions has been derived from the importance of the symmetry 
term in the equation of state and of its density dependence for 
astrophysical applications. Supernova simulations or neutron star 
models require inputs for the nuclear equation of state at extreme 
values of density and asymmetry
\cite{lattimer,lattprak,thielemann,botv04}. Isotopic effects in 
multifragmentation and other types of reactions 
have been shown to be sensitive to the symmetry energy coefficient, 
which may permit its study in the range of densities 
explored during the various stages of these collisions 
\cite{tsang01,botv02,ono03,tsang04}.

\section{EXPERIMENTAL SITUATION}

Data obtained with 4-$\pi$ detection devices have been dominating 
the field for several years \cite{nn2000,nn2003}. 
Besides solid-angle coverage, also the granularity and the dynamic 
range in particle type and energy are important parameters. 
A new standard for granularity has been set with the CHIMERA 
detector installed at the Laboratori Nazionali del Sud in 
Catania \cite{alde02}. It comprises 1192 individual Si-CsI(Tl) 
telescopes which are arranged in a ring structure providing
highest granularity at forward angles \cite{geraci04}. 

The obvious need for the coincident detection of neutrons is 
known since long ago. In heavy reaction systems, a major part of 
the released excitation energy resides in the neutron channels, 
and calorimetry without neutron detection has to rely on assumptions. 
For isotopic studies, the neutron-to-proton ratio is a primary 
observable. Measurements of neutron multiplicities and energy 
spectra in coincidence with 4-$\pi$ charged-particle detection have 
been performed with the NIMROD detector installed at Texas A\&M 
University at College Station \cite{wada04}. 
The data sets collected for selected reactions have permitted a 
comprehensive comparison with transport models.

Apart from 4-$\pi$ devices, dedicated setups of high complexity 
have been developed for specific experiments. The measurement of 
neutron-neutron and proton-proton correlation functions has 
revealed isotopic effects in the space-time properties of the 
fragmenting source at breakup \cite{ghetti04}. Fragment detection 
with mass identification is a prerequisite for studying isotopic 
effects in the fragment channels. For this purpose, new 
high-resolution arrays were constructed at Michigan State University 
\cite{wagner01} and Indiana University \cite{davin01}.

Possibilities for new experiments with existing detection systems have also 
been explored. By transporting the INDRA multi-detector \cite{pouthas} 
to GSI, a new range of energies and reactions has become accessible for
high resolution studies with 4-$\pi$ coverage. Proton-rich secondary 
beams have been employed in experiments with the ALADIN spectrometer
at GSI for the study of isotopic effects in the fragmentation of 
relativistic projectiles. First, preliminary, results from this latter 
experiment are now becoming available \cite{sfienti04}, 
some of the results obtained in the former campaign will 
be discussed further below.

\section{THEORETICAL EXPERIMENTS}

The idea of studying the equilibrium dynamics of a finite number 
of nucleons in a container has first been realized by Schnack and 
Feldmeier several years ago \cite{schnack97}. The Fermionic 
Molecular Dynamics (FMD) model was used to propagate systems 
consisting of small numbers of nucleons over sufficiently long 
times, so as to allow them to establish equilibrium. 
With a weakly coupled thermometer, the temperature was measured 
as a function of the excitation energy of the system which was 
varied. A first-order transition from a Fermi liquid to a 
Fermi vapour within the confining oscillator potential is 
clearly observed. It appears as a natural consequence of the 
short-range repulsion and long-range attraction of the
potential used for simulating the nuclear forces.

Theoretical experiments of this type, based on the 
Antisymmetrized Molecular Dynamics (AMD) model, were conducted 
by Sugawa and Horiuchi \cite{suga99} and, very recently, by 
Furuta and Ono \cite{furu03}. The outcome in these cases is very 
similar. The observed latent heat which is a function of the
chosen volume or pressure indicates a first-order transition. 
The agreement reached by these studies for the location of the 
critical temperature near 12 MeV in medium-heavy systems is remarkable.

\begin{figure}[ttb]
    \centering
    \includegraphics[width=20pc]{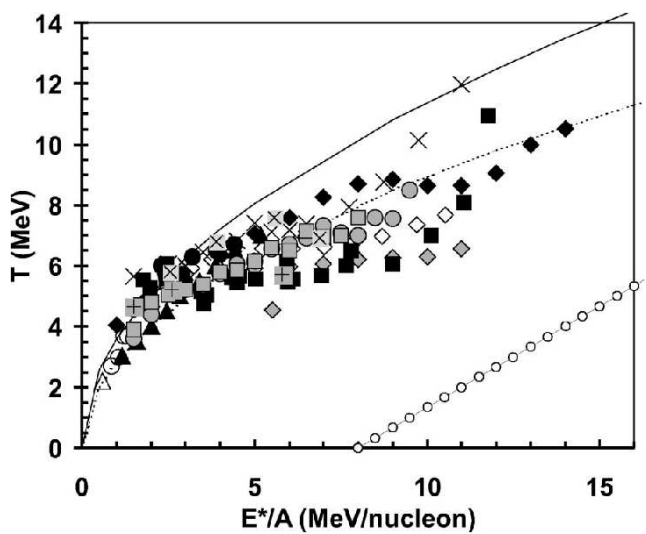} 

\vspace{-2mm}

\caption{Caloric-curve data from several experiments as compiled 
by Natowitz et al. \protect\cite{nato02}. Temperatures deduced 
from measured spectra of light charged particles and from 
double-isotope ratios of light fragments are shown as a function 
of the excitation energy deduced from the measured momentum 
transfer or from calorimetry. The Fermi-gas model predictions for
parameters $K$ = 8 and 13 are shown by the dotted and solid lines, 
respectively. The open circles represent the expectation for a 
hypothetical nucleon gas.
}

\label{fig:temp}
\end{figure}

To validate these results, it is essential that the ability of the 
FMD and AMD models of realistically describing nuclear dynamics are 
thoroughly tested, and here considerable progress has been made.
In particular, the AMD model is in the process of being intensively 
compared to reaction data, and very satisfactory agreement has been 
obtained \cite{wada04,ono02}. Both, the AMD and the FMD, have also 
been developed into successful theories for ab-initio calculations 
of the structure of light nuclei \cite{hori04,neff04}. 
The idea of deriving support for the nuclear liquid-gas phase 
transition from studies of cold nuclei seems rather intriguing.

\begin{figure}[ttb]
\centering
\begin{minipage}[t]{.45\textwidth}
    \centering
    \includegraphics[width=18pc]{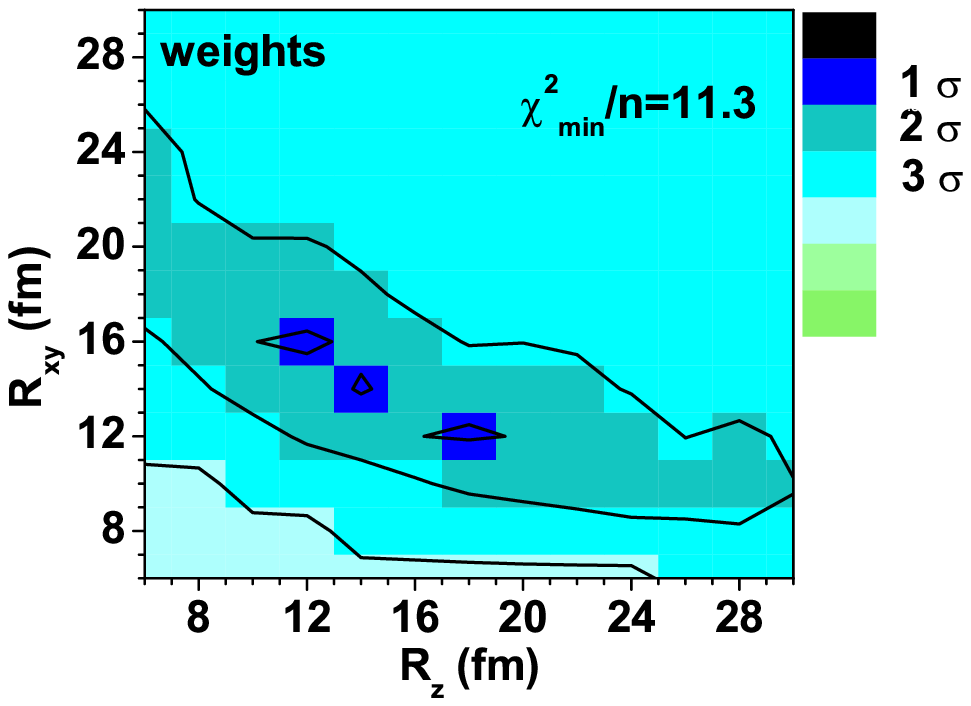} 
\end{minipage}
\hspace{\fill}
\begin{minipage}[t]{.45\textwidth}
    \centering
    \includegraphics[width=18pc]{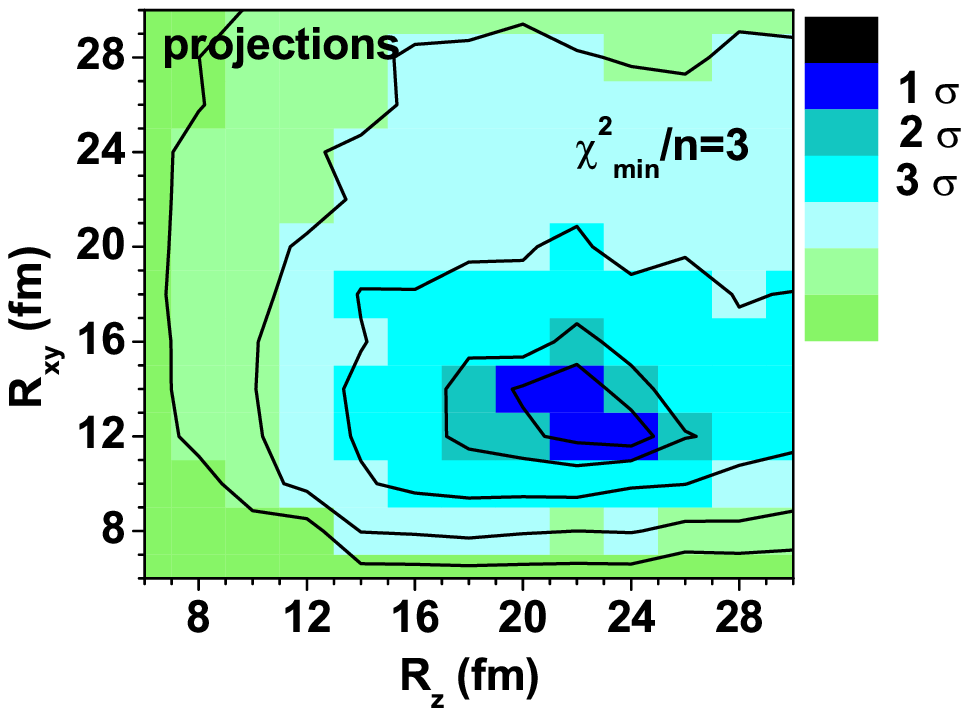}
\end{minipage}
\vspace{-5mm}

\caption{Distributions in $\chi^2$ as a function of the transverse and 
longitudinal radius of the source for weighted (left panel) and 
projected (right panel) fragment-fragment correlation functions 
($5\leq Z \leq7$) for $^{129}$Xe + $^{\rm nat}$Sn 
(from Ref. \protect\cite{lefborm03}).
}

\label{fig:chi2a}
\end{figure}

\section{THERMODYNAMICAL PARAMETERS}

Temperature measurements are useful for localizing the breakup 
process in the nuclear phase diagram \cite{poch95}. 
Here, a remarkable consistency has been reached 
recently. The systematics compiled by Natowitz et al. \cite{nato02} 
of experiments selected according to standard criteria, but 
allowing for different methods, is shown in Fig.~\ref{fig:temp}. 
The combined set of temperatures exhibit an apparent Fermi-gas-like 
rise at excitation energies up to 3-4~MeV per nucleon, a very slow 
rate of temperature increase at higher energies, and some indication 
of a further rise at the very high excitation energies above 
9~MeV per nucleon. 
Temperatures of around 6~MeV prevail in the range of 
excitation energies 5 to 8 MeV per nucleon at which 
multifragmentation is the dominant reaction channel for heavy systems.
On the scale provided by the Molecular Dynamics studies, they place 
multifragmentation well below the critical point and into the 
coexistence region where homogeneous systems cannot exist in 
equilibrium.

The apparent temperature spread (Fig.~\ref{fig:temp}) disappears 
to some part if the measurements are sorted according to the mass 
of the studied systems \cite{nato02}.
The lower temperatures in the fragmentation domain ($\approx6$~MeV) are 
associated with heavy systems of mass $A\approx200$ while the higher 
temperatures ($\approx8$~MeV) are observed for the light systems with 
$A<100$. The steady rise of temperature with excitation energy 
for the very small systems may even indicate
that here the breakup occurs close to the critical point \cite{ma04}. 
This systematic dependence has been associated with the concept 
of the limiting temperature at which excited homogeneous systems 
become unbound \cite{besp89}. According to these finite-temperature 
Hartree-Fock calculations, the limiting temperature should also 
strongly depend on the isotopic composition, a prediction to be 
tested experimentally.

The critical density of nuclear matter is approximately one 
third of the saturation density \cite{jaqa83}. 
For the statistical models of multifragmentation, the assumption 
of similarly low densities has, 
very early on, been found essential for correctly reproducing the 
observed fragment multiplicities \cite{bowman91,hubele92,deses98}. 
Many experiments have shown that
the kinetic-energy spectra of emitted fragments are inconsistent with 
emission from a normal-density compound nucleus 
\cite{milkau93,avde98,cibor00,brack04}. More direct information 
on source properties in coordinate space is accessible 
with the technique of interferometry. Here, the shape of the
source, the extension of statistical descriptions to non-spherical 
sources \cite{lefevre99}, and possible connections to transparency 
and mutual stopping in the approach phase of the collision 
\cite{reisdorf04} have received particular attention recently.

A new result obtained with a promising method of interferometry 
is shown in Fig.~\ref{fig:chi2a}. Two types of directional 
fragment-fragment correlation functions are 
used to determine the volume and the shape of the source at breakup 
following central collisions of $^{129}$Xe + $^{\rm nat}$Sn at 50~MeV per 
nucleon \cite{lefborm03}. The $\chi^2$ distributions given in the 
figure represent the quality of their fitting, under the assumption 
of negligible emission times, with results for model sources with 
different longitudinal and transverse extensions, $R_z$ and $R_{xy}$. 
Large radius parameters giving evidence for expansion are obtained 
with either technique. The correlation functions constructed from 
the projections of the relative velocities, in addition, indicate a 
longitudinally expanded shape with axis ratio $\approx 1.7$.

\begin{figure}[htb]
    \centering

    \includegraphics[width=8.5cm]{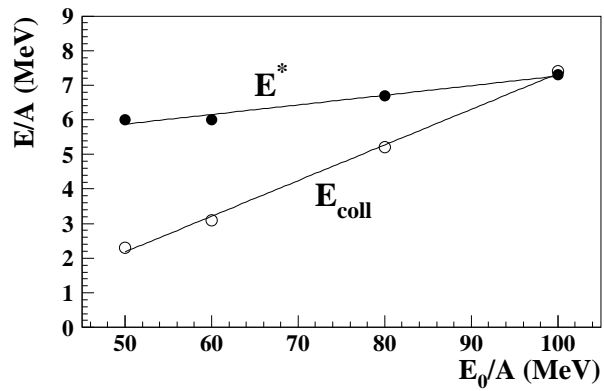} 
\vspace{-2mm}

\caption{Mean thermal excitation energy
    (full circles) and collective flow energy (open circles) at
    freeze-out, extracted by means of the MMMC-NS model,
    as a function of the incident energy $E_0/A$ for
    central collisions of $^{129}$Xe + $^{\rm nat}$Sn at 50~MeV per nucleon 
    and $^{197}$Au + $^{197}$Au at 60, 80 and  100~MeV per nucleon.
    The lines are linear fits to guide the eye
    (from Ref.~\protect\cite{lefevre04}).}

\label{fig:synth}
\end{figure}

A rapid expansion, following an initial buildup of pressure, 
will appear as a collective fragment motion. Collective radial 
flow refers to simultaneous transverse and longitudinal collective 
expansions which are observed in azimuthally inclusive experiments. 
Collective flow in central collisions of heavy systems is observed 
above a threshold energy of $\approx 50$~MeV per nucleon \cite{deses98,marie97} 
and starts to rise rapidly at higher energies. This
range of intermediate bombarding energies has been systematically 
covered in the set of experiments conducted with INDRA at GSI 
\cite{lefevre04}. The statistical analysis of these data yields 
an equilibrated excitation energy $E^*$ of the fragmenting source 
which rises rather slowly over the energy range up to 100~MeV per 
nucleon, and a rapidly rising collective-flow energy $E_{\rm coll}$. 
At 100~MeV per nucleon incident energy, both components have about 
the same magnitude (Fig.~\ref{fig:synth}).

The successful description of these highly dynamical processes with 
statistical fragmentation models, including a decoupled flow 
\cite{lefevre04}, raises the question why the effects of the 
collective motion are not more clearly visible. Collective flow 
should affect the partitioning of the system and, in particular, 
the survival probability for heavier fragments 
\cite{kunde95,pal,chikazumi,das,gulminelli,reisdorf04a}.
The effect of the flow on the charge distributions, on the other hand, 
may be simulated by adapting the value of the thermalized energy in 
the model description \cite{pal,gulminelli}.
It is, therefore, quite likely that the flow effect is implicitly 
included in the parameters of the statistical description. 
At moderate flow values, the changes are expected to be very 
small \cite{das,gulminelli}.

An alternative approach to the question of the coexistence of 
equilibrated partitions and collective motion has
recently been presented by Campi et al.~\cite{campi}.
Using classical molecular dynamics calculations and specific 
clustering algorithms, these authors find fragments to be preformed 
at the beginning of the expansion stage when the temperature and 
density are still high. The fragment charge distributions, 
reflecting the equilibrium at this early
stage when the flow is small, remain nearly unmodified down to the
freeze-out density at which the flow has fully developed.

\begin{figure}
\begin{minipage}[t]{.45\textwidth}
    \centering
    \includegraphics[height=7.0cm]{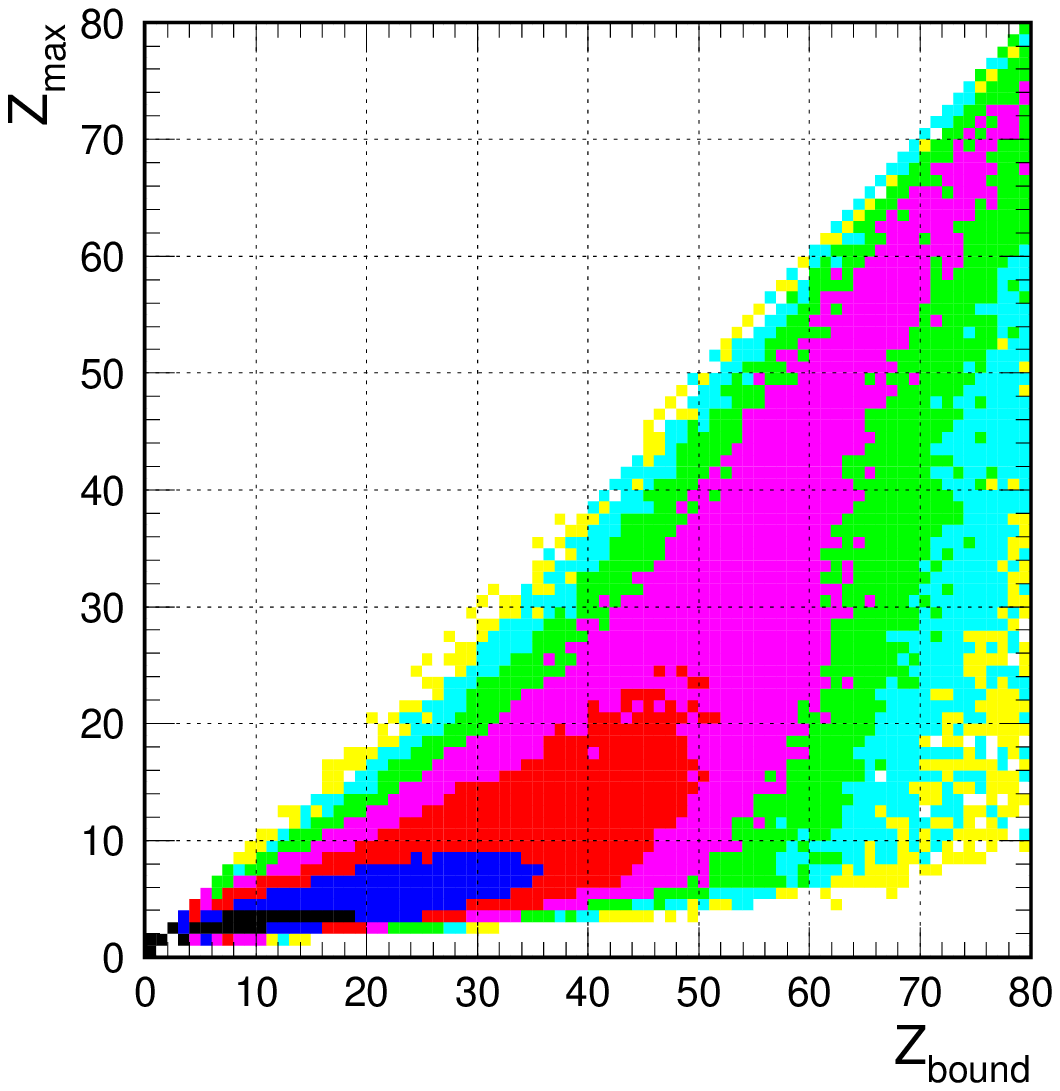} 
\vspace{-4mm}

\caption{Distribution of $Z_{\rm max}$ versus $Z_{\rm bound}$ for $^{197}$Au on 
$^{197}$Au at 1000 MeV per nucleon \protect\cite{schuett96}. 
Conventional fission events are removed . The shadings follow a 
logarithmic scale.
}

\label{fig:zbound}
\end{minipage}
\hspace{\fill}
\begin{minipage}[t]{.45\textwidth}

    \centering
    \includegraphics[height=7.0cm]{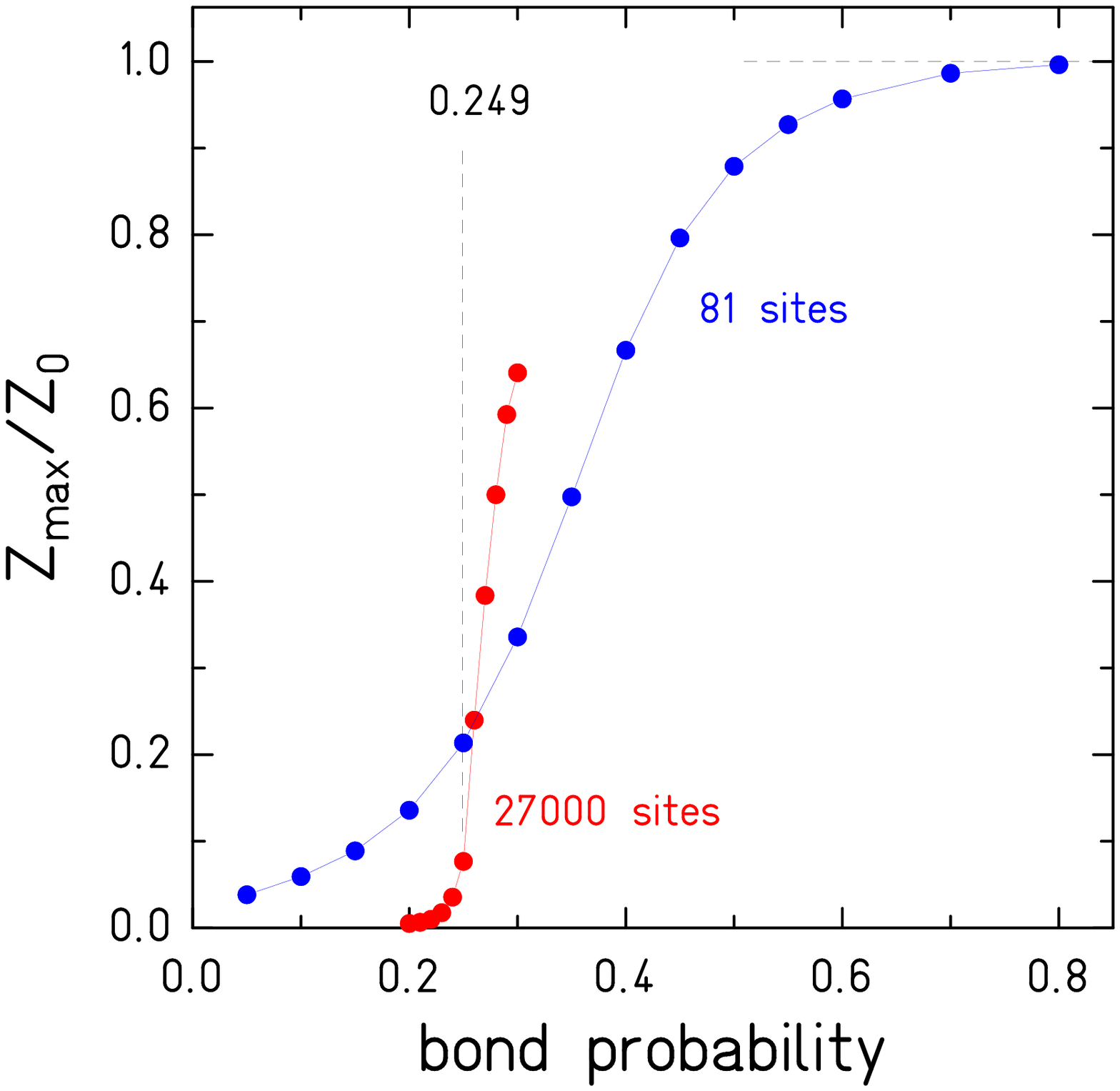} 

\vspace{-4mm}

\caption{Bond percolation: relative magnitude of the largest cluster 
as a function of the bond probability for cubic lattices of 81 and 
27000 sites. 
The critical bond probability in the infinite sytem is 0.249. 
}

\label{fig:perco}
\end{minipage}
\end{figure}

\section{LARGEST FRAGMENT AS ORDER PARAMETER}

In the search for an experimentally accessible order parameter 
of the phase transition, as observed in reactions, the largest 
fragment of the partition has appeared as a promising choice. 
It may be identified with the part of the system in the liquid phase, 
and it is correlated with the mean density which is the natural 
order parameter of a liquid-gas phase transition.

Statistical model calculations for nuclear multifragmentation
show that the disappearance of the dominating fragment is
associated with a maximum of the heat capacity which is the more 
strongly pronounced the larger the system \cite{dasgupta98}. 
The disappearance of the largest cluster, with the variation of a 
suitable control parameter, has been identified as a prominent 
signal also in fragmentations of other systems as, e.g., 
atomic hydrogen clusters \cite{gobet01}, and the largest cluster 
is an order parameter in percolation theory \cite{stauffer}. 
In finite percolation lattices, the disappearance of 
a dominant largest cluster proceeds rather smoothly and with 
obvious similarity to the nuclear experiment 
(Figs.~\ref{fig:zbound},\ref{fig:perco}).

The success of percolation models in describing the observed 
partitioning of nuclear systems
\cite{kreutz93,bauer95,kleine02} and the apparent critical behaviour 
\cite{campi88} do not necessarily identify the transition as of 
second order but rather show that first- and second-order phenomena 
may be compatible in small systems \cite{gulm99}. Studies of the 
fluctuation properties of the largest fragment \cite{frankl04} 
or of the relative magnitude of the two or three largest fragments 
\cite{pichon03} are thus not sufficient to establish 
the first-order character of the observed transition. 
Alternative methods have been presented \cite{dagostino00,elliott02}
which, however, require calorimetry on an event-by-event basis which
cannot be performed without assumptions. Also methodological aspects 
are under debate \cite{campi04,dagostino04}.  
Comparative studies of several of these signals are presently being 
performed by the INDRA collaboration \cite{lopez01,borderie04}.

\begin{figure}[htb]
\begin{minipage}[t]{.45\textwidth}
    \centering
    \includegraphics[height=8.0cm]{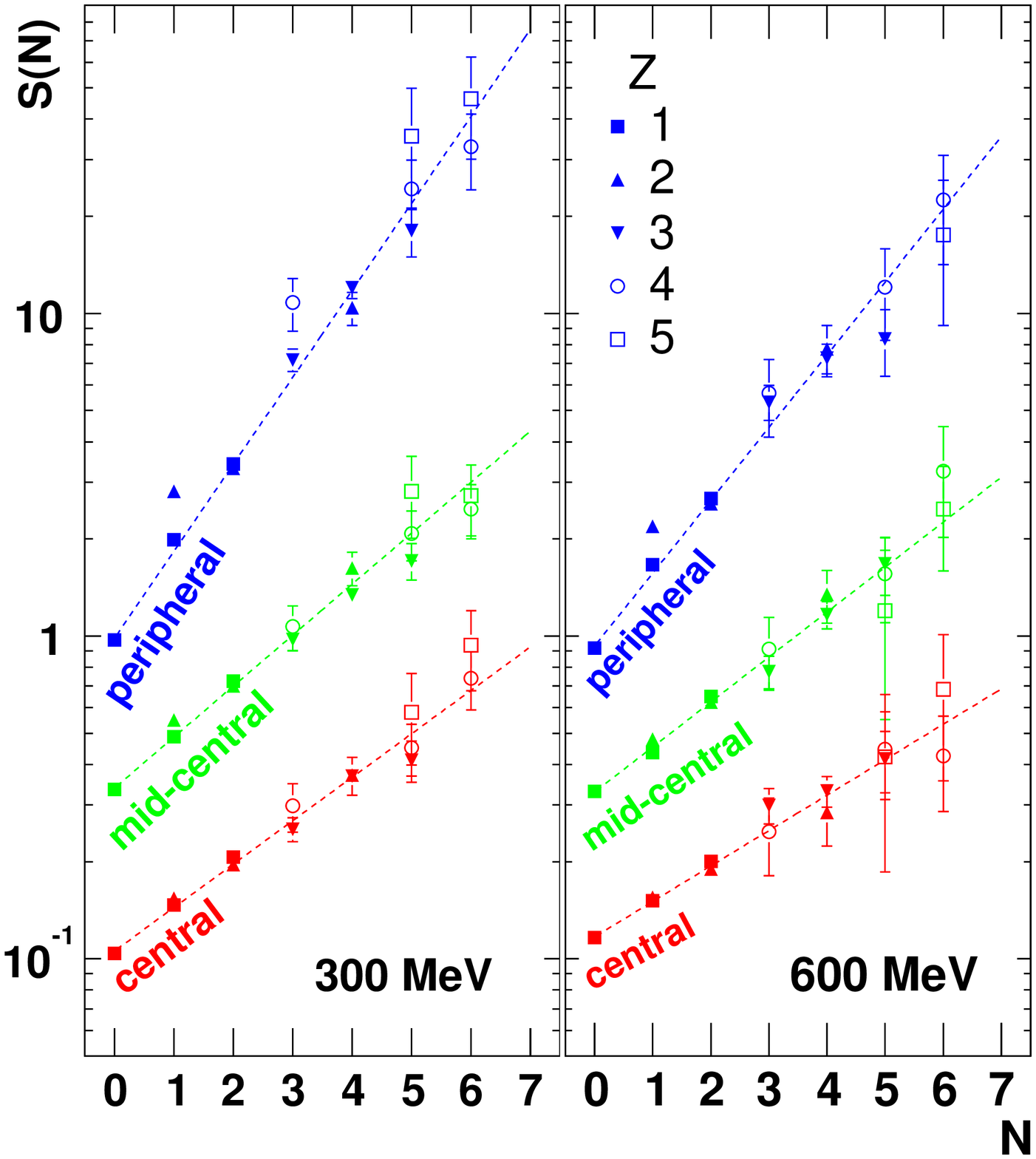} 

\vspace{-3mm}

\caption{Scaled isotopic ratios $S(N)$ for $^{12}$C + $^{112,124}$Sn at 
$E/A$ = 300~MeV and 600~MeV for intervals 
of reduced impact parameter $b/b_{\rm max}$, with
''central'' indicating $b/b_{\rm max} \leq 0.4$.
The dashed lines are the results of exponential fits according
to Eq.~(\protect\ref{eq:scalab}) (from Ref.~\protect\cite{lefsymm04}).
}
\label{fig:iso}
\end{minipage}
\hspace{\fill}
\begin{minipage}[t]{.45\textwidth}
    \centering
   \includegraphics[height=8.0cm]{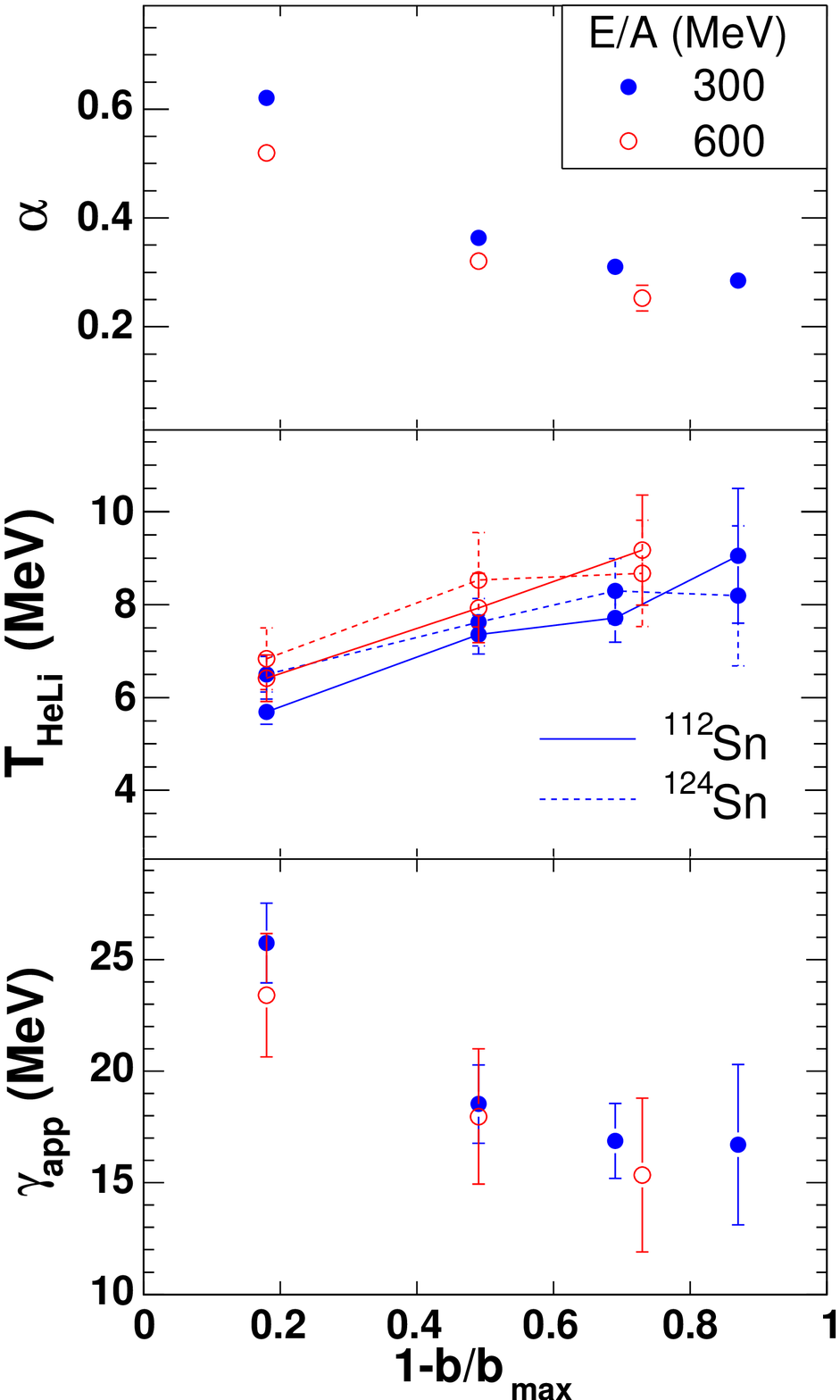} 
\vspace{-3mm}

\caption{Isoscaling coefficient $\alpha$ (top), double-isotope temperatures 
$T_{\rm HeLi}$ (middle) and resulting $\gamma_{\rm app}$ (bottom) 
for the reactions $^{12}$C + $^{112,124}$Sn at 
$E/A$ = 300~MeV (full symbols) and 600~MeV (open symbols), 
as a function of the centrality
parameter $1-b/b_{\rm max}$ (from \protect\cite{lefsymm04}).
}
\label{fig:gamma}
\end{minipage}
\end{figure}

\section{NEW DIRECTIONS}

The experimental study of particle and fragment production with 
isotopic resolution
has led to the identification of isoscaling, a phenomenon 
shown to be common to many different types of heavy ion reactions
\cite{tsang01,botv02,fried04,soul03}.
It is observed by comparing product yields
$Y_i$ from reactions which differ only in the isotopic
composition of the projectiles or targets or both. 
Isoscaling refers to an
exponential dependence of the measured yield ratios $R_{21}(N,Z)$
on the neutron number $N$ and proton number $Z$ of the detected 
products. The scaling expression
\begin{equation}
R_{21}(N,Z) = Y_2(N,Z)/Y_1(N,Z) = C \cdot exp(\alpha N + \beta Z)
\label{eq:scalab}
\end{equation}
describes rather well the measured ratios over a wide range of
complex particles and light fragments \cite{tsang01a}. 
For illustration, the scaled
isotopic ratios $S(N) = R_{21}(N,Z)/{\rm exp}(\beta Z)$ for the reactions
$^{12}$C + $^{112,124}$Sn at 300 and 600~MeV per nucleon, 
studied with INDRA at GSI \cite{lefsymm04}, are shown 
in Fig.~\ref{fig:iso}. The slope parameter $\alpha$ is found to decrease 
with increasing centrality of the reaction.

Of particular interest is the connection of the isoscaling parameters
with the symmetry-term $E_{\rm sym} = \gamma (A-2Z)^2/A$
in the nuclear equation of state which has been consistently 
established with several 
methods \cite{botv02,ono03,tsang01a}.
The coefficient $\gamma$ \cite{gamma,bond95}
is proportional to the isoscaling coefficient $\alpha$ according to
$\alpha T \approx 4\gamma \cdot (Z_{1}^2/A_{1}^2 - Z_{2}^2/A_{2}^2)$
where $T$ is the temperature and $Z_{i}$ and $A_{i}$ are the charges and 
mass numbers of the two systems at breakup. 
The results obtained for the $^{12}$C + $^{112,124}$Sn reactions are 
summarized in Fig.~\ref{fig:gamma}. Using the above equation and 
assuming that the isotopic compositions are practically
equal to those of the original targets, an apparent symmetry-term 
coefficient $\gamma_{\rm app}$ was determined, i.e. without sequential 
decay corrections for $\alpha$. The results are found to be close to 
the normal-density coefficient for peripheral collisions but drop 
to lower values at central impact parameters. If the corrections 
for sequential fragment decay after breakup are taken into account 
the resulting coefficient $\gamma$ for central collisions is even 
smaller \cite{lefsymm04}. 

This result as well as those obtained for other reactions 
\cite{tsang04,liu04} may be considered as first steps within a 
program of studying the symmetry term far from saturation density. 
In the future, such experiments may aim at profiting 
from the existence of new radioactive-beam facilities.
Projectiles with large neutron excess seem to be best 
suited to study dynamical effects of the symmetry force, 
and central collisions at high incident energy will have to be 
chosen in order to reach the larger than normal densities at which the 
symmetry energy is least well known \cite{bao02,greco02}.

Stimulating discussions with my colleagues at the GSI and within the
ALADIN and INDRA-ALADIN Collaborations are gratefully acknowledged.
In addition, I would like to thank R.T.~de Souza, H.~Feldmeier, 
T.~Furuta, J.B.~Natowitz, H.~Oeschler, and A.~Pagano for
providing me with graphics for the talk and for helpful comments.


\begin{thebibliography}{9}
\bibitem{gamma}
the notation follows that originally chosen in Ref. \protect\cite{bond95};
alternatively $C_{\rm sym}$ is frequently used for the same quantity.
\end{thebibliography}
\end{document}